\documentclass[sigconf]{acmart}
\acmConference[ICPC 2022]{The 30th International Conference on Program Comprehension}{May 21–22, 2022}{Pittsburgh, PA, USA}
\usepackage{booktabs}
\usepackage{makecell}
\usepackage{multirow}
\usepackage{listings}
\usepackage{xcolor}
\usepackage{graphicx}
\usepackage{subcaption}
\usepackage{pgfplots}
\usepackage{framed}
\pgfplotsset{compat=newest}

\newcommand{\ie}{{\textit{i.e.}}, }
\newcommand{\eg}{{\textit{e.g.}}, }

\definecolor{codegreen}{rgb}{0,0.6,0}
\definecolor{codegray}{rgb}{0.5,0.5,0.5}
\definecolor{codepurple}{rgb}{0.58,0,0.82}
\definecolor{backcolour}{rgb}{0.95,0.95,0.92}

\lstdefinestyle{mystyle}{
  commentstyle=\color{codegreen},
  keywordstyle=\color{magenta},
  numberstyle=\tiny\color{codegray},
  stringstyle=\color{codepurple},
  basicstyle=\footnotesize,
  breakatwhitespace=false,         
  breaklines=true,                 
  captionpos=b,                    
  keepspaces=true,                 
  numbers=left,                    
  numbersep=5pt,                  
  showspaces=false,                
  showstringspaces=false,
  showtabs=false,                  
  tabsize=2,
  frame = single
}

\AtBeginDocument{%
  \providecommand\BibTeX{{%
    \normalfont B\kern-0.5em{\scshape i\kern-0.25em b}\kern-0.8em\TeX}}}

\copyrightyear{2022} 
\acmYear{2022} 
\setcopyright{acmcopyright}\acmConference[ICPC '22]{30th International Conference on Program Comprehension}{May 16--17, 2022}{Virtual Event, USA}
\acmBooktitle{30th International Conference on Program Comprehension (ICPC '22), May 16--17, 2022, Virtual Event, USA}
\acmPrice{15.00}
\acmDOI{10.1145/3524610.3527880}
\acmISBN{978-1-4503-9298-3/22/05}

\begin{document}

\title{Clone-based code method usage pattern mining}
\author{Zhipeng Xue}
\authornote{Corresponding author}
\email{xuezhipeng19@nudt.edu.cn}
\affiliation{%
  \institution{National University of Defense Technology}
  \city{Changsha}
  \state{Hunan}
  \country{China}
}

\author{Yuanliang Zhang}
\email{zhangyuanliang13@nudt.edu.cn}
\authornote{Co-first author}
\affiliation{%
  \institution{National University of Defense Technology}
  \city{Changsha}
  \state{Hunan}
  \country{China}
}

\author{Rulin Xu}
\email{xurulin11@nudt.edu.cn}
\affiliation{%
  \institution{National University of Defense Technology}
  \city{Changsha}
  \state{Hunan}
  \country{China}
}

\begin{abstract}

When programmers retrieve a code method and want to reuse it, they need to understand the usage patterns of the retrieved method. 
However, it is difficult to obtain usage information of the retrieved method since this method may only have a brief comment and few available usage examples. In this paper, we propose an approach, called LUPIN (cLone-based Usage Pattern mIniNg), to mine the usage patterns of these methods, which do not widely appeared in the code repository. The key idea of LUPIN is that the cloned code of the target method may have a similar usage pattern, and we can collect more usage information of the target method from cloned code usage examples. From the amplified usage examples, we mine the usage pattern of the target method by frequent subsequence mining after program slicing and code normalization. Our evaluation shows that LUPIN can mine four categories of usage patterns with an average precision of 0.65.

\end{abstract}


\maketitle
\section{Introduction}
Programmers sometimes need to use some methods they are not familiar with, such as methods retrieved from code search tools. However, understanding the usage of methods will require considerable time \cite{hansen2013makes}, or it could lead to latent bugs \cite{jiang2007context,li2006cp}. To avoid misusing such methods, programmers often refer to code comments \cite{lee2018comment} (e.g., JavaDoc) or search in question and answer (Q \& A) forums \cite{uddin2020mining,zhang2018code} (e.g., Stack Overflow). Unfortunately, most method comments only have insufficient information, which only describes the functionality of the methods instead of the usage patterns or constraints of methods \cite{haouari2011good}. Moreover, the Q \& A forums have little information about a method if it is not a widely used application programming interface (API) or in a project with high popularity.

To help programmers use the code method correctly, previous studies try to mine the usage of the code method. MUSE \cite{10.5555/2818754.2818860} collected all the usage examples of the given method and demonstrated the best usage examples of given methods according to a popularity ranking. CodeKernel \cite{gu2019codekernel} built object usage graphs for all the usage examples and outputted a representative example by graph kernel. Although these tools can provide several useful and reliable usage examples, programmers still need to summarize the usage patterns or constraints from usage examples manually. Ren \textit{et al.} \cite{ren2020api} mined the usage pattern of APIs and detected API misuse by building a knowledge graph from documents. UP-Miner \cite{wang2013mining} mined the API usage pattern by clustering frequent closed sequences. However, these studies mainly focus on APIs, which have complete documents or sufficient usage examples.

In this paper, we focus on mining the usage patterns of code methods that may have insufficient manuals to refer to and only have few usage examples. Since it is difficult to mine correct usage patterns from insufficient usage examples, we obtain more reliable usage information with the help of the code clone. The core idea of our approach, called LUPIN (cLone-based Usage Pattern mIniNg), is that the cloned code of the target method may have similar usage patterns, and we can collect more usage information of the target method from cloned code usage examples. Based on the usage examples of the target method and its cloned method, we can mine its usage pattern by frequent subsequence mining. To the best of our knowledge, LUPIN is the first tool to mine usage patterns of code methods from their cloned code.

This paper addressed two important challenges, as shown below. First, since some statements in the usage example are irrelevant to the target method, we need to filter out these statements to reduce the cost of frequent subsequence mining. Second, different programming styles of the same functionality (\eg {\tt for} and {\tt while}) may be regarded as different subsequences, which leads to the omission of reported results.

To solve these challenges, we first perform program slicing to filter out irrelevant statements. The strict program slicing strategy will lose useful information, while the relaxed program slicing strategy will reserve noise information. We leverage program slicing based on the program dependency graph (PDG) to retain dataflow-related statements. To avoid disruptions from different programming styles, we normalize the usage example code at the variable, statement, and syntax levels.

To evaluate the performance of LUPIN, we employ it on a Java project dataset 50k-c \cite{martins201850k}. From dataset 50k-c, we collect 11,414 methods that have more than 10 separate usage examples. From the usage examples of target methods and their cloned methods, we mine 2,368 usage patterns. The mined usage patterns can be classified into 4 types based on syntax keywords: condition check, iteration, error handling, and method co-occurrence. According to a manually check in the subset of mined usage patterns, the average precision of LUPIN is 0.65.
\section{Motivation Example}
\label{motivation}

\begin{figure}[htbp]
    \centering
    \begin{subfigure}{0.45\textwidth}
        \centering
        \begin{lstlisting}[language=java]
...
dot = writeDotSourceToFile(dotSource);
if (dot != null) {//condition check
    img_stream = get_img_stream(dot, type, representationType);
    if (!dot.delete()) {
    ...
    }
}
      \end{lstlisting}
      \subcaption{The code snippet 1}
      \label{motivation code a}
    \end{subfigure}
    \begin{subfigure}{0.45\textwidth}
        \centering
        \begin{lstlisting}[language=java]
...
dot = writeDotSourceToFile(dotSource);
if (dot != null){//condition check
   img_stream = get_img_stream(dot, type, "dot", dpi);
}
...
if(dot!=null) {
    dot.delete();
}
\end{lstlisting}
      \subcaption{The code snippet 2}
      \label{motivation code a}
    \end{subfigure}
    \caption{Usage Examples of a Method and its Cloned Method}
    \label{motivation code}
  \end{figure}
In this section, we first give two code snippets to illustrate that the target method and cloned method have similar usage patterns.

{\tt get\_img\_stream} method is used in more than 50 projects in Github, while there is no clear and unambiguous comment or document about its usage pattern. The code snippet 1 in Fig.\ref{motivation code} is a usage example of the {\tt get\_img\_stream} method in {\tt confluence-graphviz} project and code snippet 2 in Fig.\ref{motivation code} is a usage example of its cloned method in {\tt JavaCommExtractor} project. Although the target method {\tt get\_img\_stream} only has three arguments while the cloned method has four arguments instead, they have similar functionality and usage patterns. The method {\tt get\_img\_stream} calls the external dot program and returns the image in binary format. From both code snippets, we can find that the programs should check the value of the first parameter of {\tt get\_img\_stream}(\ie {\tt dot}). If {\tt dot} is {\tt NULL}, the programs will not execute {\tt get\_img\_stream}. However, if programmers use {\tt get\_img\_stream} while missing the condition check, it will lead to an exception termination.

The same usage information from the target method and its cloned method inspires us to search the usage pattern of the target method not only from its usage examples but also those of its cloned methods.
\section{Proposed Method}
\label{proposed method}

  \begin{figure}[htbp]
	\centering
	\includegraphics[width=8cm]{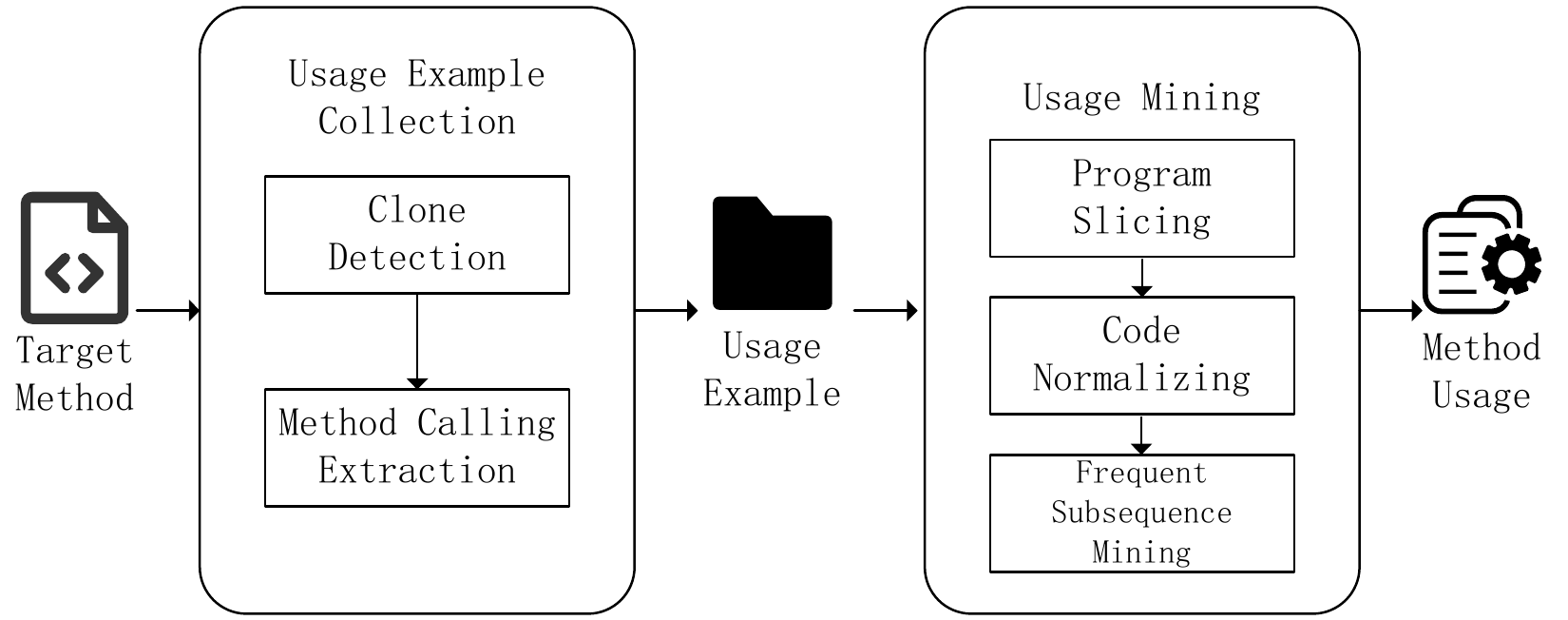}
	\caption{The Overview of LUPIN}
	\label{overview}
\end{figure}

In this section, we propose LUPIN (cLone-based Usage Pattern mIniNg) to mine the usage pattern of a code method and guide programmers on how to use it correctly.

\subsection{Overview}
Fig.\ref{overview} illustrates the overview of LUPIN. There are two major components to LUPIN. The first component detects the cloned code of the target method and collects all the usage examples of the target method and its cloned methods. The second component mines usage patterns from collected usage examples. It applies program slicing to remove useless code, normalizes the relevant code, and mines the usage patterns by frequent subsequence mining for the target method.

\subsection{Usage Example Collection}
Firstly, we need to collect the cloned methods of target method and their usage examples. To avoid the false positives of code clones, we only collect Type-1 and Type-2 code clone \cite{roy2009comparison}. We leverage \textit{CloneWorks} \cite{svajlenko2017cloneworks}, which is one of the state-of-the-art tools to handle fast and large-scale clone detection with over 80\% precision and recall.

To collect the usage examples of the target method and its cloned method, we transform each method in the code repository into an AST by \textit{Javalang\footnote{https://github.com/c2nes/javalang}}, and then we look for the method calls by visiting every AST node. If the AST node calls the target method or its cloned methods, the corresponding code will be collected as a usage example of cloned methods.

\subsection{Program Slicing}
To filter out the statements irrelevant to the called method of interest in usage examples, we employ program slicing \cite{weiser1984program}. In this paper, we slice the program based on a program dependency graph (PDG). Usually, the PDG is generated according to the information during compilation, while it will spend expensive time and space compiling all the projects in the code repository. We leverage a tool \textit{JCoffee} \cite{gupta2020jcoffee} to complement each usage example code and make it compliable. Based on the PDG of usage example code, we use both forward and backward program slicing and retain only the statements that connect to the called method by the data and control flow edge directly. 

\subsection{Code Normalizing}
Since the code programming style is variable, separate codes may perform functional similarity. For example, {\tt if(a\textgreater b)} and {\tt if(b\textless a)} indicates the same condition, they will be regarded as two different elements in the frequent subsequence mining algorithm. Therefore, it is necessary to normalize the code and reduce the noise from the code programming style. In this paper, we normalize the code at the variable level, statement level, and syntax level.

\textbf{Variable-level normalizing.}
Since there is no mandatory rule for variables, the names of variables can vary greatly. To reduce the noise introduced from the variable name, we normalize the variable name by the following rules. Fig.\ref{variable normalizing} illustrates the code of Fig.\ref{motivation code}(a) after variable-level normalizing.

For the parameter of the called method, we rename them and their context by {\tt arg0, arg1...}. For example, the parameter {\tt dot, type, representationType} in Fig. \ref{motivation code}(a) is replaced by {\tt arg0, arg1, arg2} in Fig. \ref{variable normalizing}.

For the variable that indicates the return value of the called method, we rename it by {\tt ref} in the called method and its context. For example, the variable img\_stream in Fig. \ref{motivation code}(a) is replaced by {\tt ref} in Fig. \ref{variable normalizing}.

For the remaining variables in the code, we replace them by their data types. For example, the variable {\tt dot\_source} in Fig.\ref{motivation code}(a) is replaced by {\tt String} in Fig. \ref{variable normalizing}.

\begin{figure}[htbp]
        \centering
        \begin{lstlisting}[language=java]
...
arg0 = writeDotSourceToFile(String);
if (arg0 != null) {
    ref = get_img_stream(arg0, arg1, arg2);
    if (!arg0.delete()) {}
...
      \end{lstlisting}
      \caption{Example of variable-level normalizing}
    \label{variable normalizing}
  \end{figure}

\textbf{Statement-level normalizing.}
Since the different forms of statements may maintain semantic similarity, we perform statement-level normalization to reduce the inconsistencies. In this paper, we mainly focus on arithmetic operations and relation operations.

Commutative relational operations (\eg "+", "*", "!=", "\&\&") refer to operations whose functionality will not change if two operands are exchanged. For the two operands of the operation, we normalize them by lexicographical order. For example, we transform {\tt int + char} into {\tt char + int}. Since {\tt char} is lexicographically before {\tt int}.

Noncommutative relational operations (\eg "/", "\textgreater=") are opposite to commutative relational operations and will change the functionality if two operands are exchanged. For the two operands of the operation, we normalize them by lexicographical order. If we change the order of operands in the original code, we then replace the operator with its opposite operator. For example, we transform {\tt int\textless char} into {\tt char\textgreater int}. Since {\tt char} is lexicographically before {\tt int}, operator {\tt \textless} is replaced by operator {\tt \textgreater}

\textbf{Syntax-level normalizing.}
The same semantic code may be implemented in separate programming syntaxes. For example, both {\tt if-else} and {\tt switch-case} can represent the same conditional branch, while they are written in different syntactic structures. To reduce the noise from syntax, we perform syntax-level normalizing.

Since {\tt while}, {\tt for}, {\tt do-while} can represent loop structure, we normalize all the loop structures into {\tt while}. The syntax of {\tt for} structure is {\tt for(initialization; condition; increment) \{statements;\}}, then we transform it in {\tt while} structure like {\tt initialization; while(condition) \{statements; increment\}}.

The conditional branch can be implemented by {\tt if-else} and {\tt switch-case}, and we normalize all the conditional branches in {\tt if-else}. The syntax of the {\tt switch-case} structure is {\tt switch(exp)\{ case value1: statement1; break; case value2: statement2; break; default statement3;\}}. We transform it into an {\tt if-else} structure such as {\tt if(exp == value1) statement1; otherwise, if(exp == value2) statement2; otherwise statement3}.

\subsection{Frequent Subsequence Mining}
Based on the result of code normalization, we use frequent subsequence mining to summarize the target method usage pattern. Frequent subsequence mining can investigate the most common statements in all usage examples, which may reveal the usage pattern of the target method.

The usage examples of the target method and its cloned methods can constitute a code cluster $C$. The normalized usage example codes can be regarded as normalized statement sequences. We extract statements subsequence $s$ from normalized statements sequences and compute the $Support$ of each subsequence by Equation \ref{support} \cite{zaki2001spade}.
  \begin{align}
  \label{support}
  Support(s) &= \frac{\#\; of\;statement\;sequence\;including\;s}{ \#\;of\;statement\;sequence\;in\;C}
\end{align}
The subsequence whose $Support$ is greater than a threshold $\sigma$ is labeled the usage pattern of the target method. For a code method, we check whether the mined usage pattern is a subset of others, and then we discard all the subset usage patterns. In this paper, we empirically set the threshold $\sigma$ to 0.9.
\section{Evaluation}
\label{experiment}

To evaluate the method usage pattern mining performance of LUPIN, we evaluate it on the 50k-c dataset \cite{martins201850k}. The 50k-c dataset includes 50K JAVA projects that are collected from the online code repository GitHub. From the 50k-c dataset, we collect 93,587 methods that have cloned methods. Since the usage patterns mined from a few usage examples are unreliable, we filter out the methods with fewer than 10 usage examples. Then, we reserve 11,414 methods with more than 10 different usage examples. In the last, LUPIN reports 2,368 usage patterns.

According to the syntax keyword in mined usage patterns, we classify the usage patterns into four categories: condition check, iteration, error handling, and method co-occurrence. The details of the usage pattern categories are listed as follows:

\textbf{Condition Check.} Usually, condition check refers to the constraint that needs to be satisfied before or after executing the method. The missing condition check will lead to unexpected behavior of the program, such as crashing or the returning of the wrong value. Fig.\ref{motivation code} is a typical example of condition check. The cloned method {\tt get\_img\_stream} will read the image from the address pointed by parameter {\tt dot}. If a programmer misses the condition check of parameter {\tt dot}, the null pointer of {\tt dot} will trigger a IOException during executing cloned method {\tt get\_img\_stream}. 

\textbf{Iteration.} Some cloned methods should be executed in the loop and check the condition repeatedly. The missing iteration issue of cloned method usage will lead to incomplete method execution, and perform unexpected results. Fig.\ref{iteration} is an example of iteration usage of the cloned method. A programmer reads a CSV file by reusing the cloned method {\tt readnext}. The cloned method {\tt readnext} only read a line of a CSV file and checks if the CSV file has the next line. If invoke cloned method {\tt readnext} without iteration, it will only read a line of the file instead of the whole file.

 \begin{figure}[htbp]
    \centering
        \centering
        \begin{lstlisting}[language=java]
//RepSys/src/util/csvParser/CSVReader.java
public List<String[]> readAll() throws IOException {
    List<String[]> allElements = new ArrayList<String[]>();
    while (hasNext) {//iteration
        String[] nextLineAsTokens = readNext();
        if (nextLineAsTokens != null)
            allElements.add(nextLineAsTokens);
    }
    return allElements;
}
      \end{lstlisting}
    \caption{The example of iteration usage}
    \label{iteration}
  \end{figure}

\textbf{Error handling.} Error handling refers to the routines that respond to an unexpected state of cloned method. Fig.\ref{error handling} shows the example of reusing cloned method {\tt encodeBytes}. If there is something wrong when encoding the source file, the program will throw an IOException and collect the message of IOException.

 \begin{figure}[htbp]
    \centering
        \centering
        \begin{lstlisting}[language=java]
//Turkit/src/main/java/edu/mit/csail/uid/turkit/util/Base64.java  
public static String encodeBytes( byte[] source ) {
    try {
        String encoded = encodeBytes(source, 0, source.length, NO_OPTIONS);
    } catch (java.io.IOException ex) {//error handling
        assert false : ex.getMessage();
    }   
    return encoded;
}   
      \end{lstlisting}
    \caption{The example of error handling}
    \label{error handling}
  \end{figure}

\textbf{Method Co-occurrence.} Some methods should be used together, such as open files and close files. Missing one of them will lead to a memory leak or visiting an invalid source. Fig.\ref{occurrence} gives an example of cloned method co-occurrence detected by PDG-based program slicing. {\tt openBrace} and {\tt closeBrase} are a pair of cloned method, {\tt openBrace} print an opening brace and {\tt closeBrase} print a closing brace. Missing one of the cloned methods will cause inconsistency of the brace.

 \begin{figure}[htbp]
    \centering
        \centering
        \begin{lstlisting}[language=java]
//deps/jode/flow/SynchronizedBlock.java	
public void dumpInstruction(TabbedPrintWriter writer){
		writer.openBrace();//method co-occurrence
		bodyBlock.dumpSource(writer);
		writer.closeBrace();//method co-occurrence
	}
      \end{lstlisting}
    \caption{The example of method co-occurrence}
    \label{occurrence}
  \end{figure}

Since it is hard to check the correctness of mined usage patterns, similar to previous studies \cite{zhong2009mapo, zhang2018code}, we check a subset result of LUPIN manually. For each category, we randomly select 100 methods and their corresponding mined usage patterns. For a method, if its usage example will get an unexpected result or lead to abnormal termination after deleting the mined usage patterns, it indicates the mined usage patterns are necessary and correct. The result of LUPIN is shown in Table \ref{result}.

\begin{table}[htbp]  
\centering
   \caption{\label{result}Result of LUPIN in 50k-c Dataset}  
    \begin{tabular}{c|cc}  
     \toprule 
     Usage category & \# of mined usage & Precision\\
       \hline
    Condition check & 681 & 0.77 \\
     Iteration & 165 & 0.38 \\
     Error handling & 1043 & 0.82\\
     Method co-occurrence & 479 & 0.24\\
       \bottomrule
    \end{tabular} 
 \end{table}
 
 According to Table.\ref{result}, the average precision of LUPIN is 0.65. Error handling is the most usage pattern with the highest precision. Since most error handlings only print the exception or the trace information which is a necessary and popular behavior. Method co-occurrence performs the lowest precision. the most false positives are caused by irrelevant operations on the same object. For example, there is no relationship between method {\tt SelectDate} and {\tt PrintData} in database project, while they often present together.

\section{related work}

When programmers determine a required code method and want to reuse it, they need to understand the usage of the code method. There is much work focusing on usage mining, and the usages mined by previous studies mainly include usage examples \cite{10.5555/2818754.2818860, gu2019codekernel,radevski2016towards}, method call graphs \cite{mcmillan2013portfolio,nguyen2009graph} and usage patterns \cite{nguyen2019focus,zhong2009mapo}. The first two mined usages can be applied to any code method, but programmers still need to analyze the usage patterns or constraints of the methods manually. The usage pattern can be more intuitive for programmers, while most current studies mined usage patterns only for APIs, which have clear documents and sufficient usage examples. In contrast, LUPIN can mine the usage pattern for code methods which only have insufficient manuals or few usage examples.
\section{Future Work}
\label{conclusion}


In the future, we plan to extend this work from three perspectives:

\textbf{Improving LUPIN.} The simply program slicing is not sufficient for LUPIN and lead to the low precision of LUPIN especially when detect iteration and method co-occurrence usage pattern. We plan to analyse the different feature between true-positive and false-positives, and build a filter by program analyse based on the analysed feature.

\textbf{Building a benchmark of code usage patterns.} The benchmark consists of the code methods and their usage patterns. With such benchmark, we can calculate both precision and recall of LUPIN without manually effort.

\textbf{Doing a user study.} To evaluate the usefulness of mined usage patterns, we plan to do a user study to assess how useful are the recommended usage patterns to the developers.

\bibliographystyle{ACM-Reference-Format}
\bibliography{references}

\end{document}